\documentclass[preprint,aps, preprintnumbers, nofootinbib]{revtex4}
\usepackage[dvips]{graphics}

\newcommand{\lsim}[1]{
\setlength{\unitlength}{12pt}
\begin{picture}(1.4,1.)
\put(.7,-0.3){\makebox(0.0,1.)[t]{$<$}}
\put(.7,-0.3){\makebox(0.0,1.)[b]{$\sim$}}
\end{picture}#1}

\newcommand{\gsim}[2]{
\setlength{\unitlength}{12pt}
\begin{picture}(1.4,1.)
\put(.7,-0.3){\makebox(0.0,1.)[t]{$>$}}
\put(.7,-0.3){\makebox(0.0,1.)[b]{$\sim$}}
\end{picture}#2}

\begin{document}

\title{Holographic constraint and effective field theories with N-species}

\author{R. Horvat}
\email{horvat@lei3.irb.hr}
\address{Rudjer Bo\v{s}kovi\'{c} Institute, P.O.B. 180, 10002 Zagreb,
Croatia}

\begin{abstract}
Effective field theories that manifest UV/IR mode mixing in such a way as to
be valid for arbitrarily large volumes, can be used for gravitational,
non-black hole events to be accounted for. In formulating such theories with
a large number of particle species $N$, we employ constraints from the muon
$g-2$, higher-dimensional operator corrections due to the required UV and IR
cutoffs as well as the RG evolution in a conventional field-theoretical model 
in curved space. While in general our bounds on $N$ do reflect $N
\simeq 10^{32}$, a bound motivated by the solution to the hierarchy problem in the
alike theories and obtained by the fact that strong gravity has not been
seen in 
the particle collisions, the bound from the muon $g-2$ turns out to be much
stronger, $N \lsim 10^{19}$. For systems on the verge of
gravitational collapse, this bound on $N$ is far too restrictive to allow 
populating a large gap in entropy between those systems and that of
black holes of the same size.    
\end{abstract}

%\pacs{97.60.Gb, 14.60 Pq, 04.80.Cc}
\newpage

\maketitle

For a system on the brink of experiencing a sudden collapse to a black hole,
$L \sim R_s$, where $L$ is the size of the region (providing an IR cutoff) 
and $R_s$ is the gravitational
radius, a serious problem does arise in that its entropy scales as
$L^{3/2}M_{Pl}^{3/2}$, compared to that of black holes, $L^2M_{Pl}^2$ (see,
e.g. \cite{1}). If the system under consideration is the whole universe at
the preset time, the gap in entropy between those systems would be as huge
as $10^{122}$. To approach the problem, we have developed \cite{2} 
a large-$N$
formulation of the  effective quantum field theory (QFT) with UV/IR mixing 
originally derived in \cite{3}.

Previous considerations for the maximum possible entropy
suggest that ordinary QFT may not be valid for arbitrarily large volumes,
unless the UV and the IR cutoffs obey a constraint, $L \Lambda^3 \lsim
M_{Pl}^2$ \cite{3}. In view of the fact that an effective QFT
is unlikely to provide an adequate description of any system 
containing black holes \cite{4, 5}, the above constraint proclaims too weak
and should be replaced by a far more restrictive one, $L^{3/2} \Lambda^3
\lsim M_{Pl}^{3/2}$ \cite{3}. Thus, with such a constraint the holographic
information is manifestly  
encoded in an ordinary QFT, meaning that the theory not
only regains authenticity in arbitrarily large volumes, but also its
application to particle phenomenology would inevitably capture some
gravitational phenomena as well.       

With a window to quantum gravity set in, the above field-theoretical
framework has recently gained extra popularity as an original dark-energy
approach, generically dubbed `holographic dark energy' \cite{6}. The
main reason of why the above dark-energy model is so appealing in possible 
description
of dark energy of the universe is  when the holographic bound is saturated,
$L^{3/2} \Lambda^3
\simeq M_{Pl}^{3/2}$, the energy density $\Lambda^4$ gives the right amount of dark energy in the
universe at present, provided $L$ today is of order of the Hubble parameter.
Moreover, since $\Lambda^4$ is now a running quantity, it also
has a  potential  to shed some light on the `cosmic coincidence problem'
\cite{7}. In addition, the original model \cite{3} is capable to satisfy
current observations \cite{8}, since by construction it has
the equation of state $\omega_{\Lambda } = -1$. 
 
With the introduction of a large number of particle species $N >> 1$ in the
theory, we have shown \cite{2} that the saturated version of the above
constraint generalizes to 
\begin{equation}
N \Lambda^4 L^4 \simeq L^2 M_{PL}^2 \;,
\end{equation}
which means that the upper bound on the QFT entropy $N \Lambda^3 L^3$, unlike
the total energy, now 
becomes $N$-dependent,
i.e.,
\begin{equation}
N \Lambda^3 L^3 \simeq N^{1/4}  L^{3/2} M_{PL}^{3/2} \;.
\end{equation}
The internal consistency of the theory then  requires the upper bound on
$N$ to be 
\begin{equation}
N \lsim L^2 M_{PL}^2 \;,
\end{equation}
which implies that a normal system having the entropy $N \Lambda^3 L^3$, begins
to sustain a black hole entropy $L^2 M_{PL}^2$ ($N$-independent) at saturation.  

At saturation, our bound (3) becomes closely related to 
the bound on the gravitational cutoff of the (four dimensional) theory
$\Lambda = M_{PL}/\sqrt{N}$, obtained in the scenario \cite{9}. This bound
was supported its claim with a number of different arguments including the 
black hole
evaporation argument \cite{10} as well as  with considerations from 
quantum information theory \cite{11}. The hierarchy between the Planck scale
and the weak scale can then be accounted for with $N \simeq 10^{32}$. The same
number also characterizes the higher-dimensional scenario \cite{12}, hosting
additional particles of the Kaluza-Klein type.  Indeed, (3) at saturation,
together with (1) or (2), yields $\Lambda \simeq L^{-1}$, giving therefore
$\Lambda \simeq M_{PL}/\sqrt{N}$. The breakdown of an effective QFT 
with $\Lambda \simeq L^{-1}$ clearly shows a non-perturbative
nature of this bound \footnote{Unlike the present case, the correspondence
$\Lambda \simeq L^{-1}$ in the AdS/CFT context means correspondence
between the two different theories.}. Here it is just a possibility to 
downside 
$\Lambda$ much below
$\Lambda \simeq M_{PL}/\sqrt{N}$ that permits us to address the cosmological 
constant problem \cite{3}. Consequently this also weakens the bound on $N$, as seen
from the study of
the early-universe evolution, with a choice for $L$ in the form of
the future event horizon \cite{2}. The 
bound obtained there, $N \lsim 10^{68}$, as model-dependent, can be
improved by a few orders of magnitude, but not more.

In the present note, we show that  constraints from particle phenomenology 
on the number of species may be so severe that the saturation in (3) is 
actually never approached. In particular, high precision experiments would
constraint $N$ to lay much below the typical value $N \simeq 10^{32}$
from the  
alike theories. In this case consequently it is no longer possible that 
the gap in entropy 
between black hole and non-black hole configurations be completed by
additional species $N >> 1$.       

We begin our considerations with the constraint from the muon $g - 2$. Here  
we take up the same line
of reasoning as already used in \cite{3} for the $N \simeq 1$ case. Of the
essence here is to notice that a contribution from radiative corrections which
otherwise would tend to zero when both $\Lambda, L \rightarrow \infty$, now 
yields a finite answer because $\Lambda$ and $L$ 
should obey a
nontrivial constraint   from the UV/IR
mixing. Although not a usual Planck-scale correction, this contribution is
arguably of the gravitational origin, and should be regarded as a correction
beyond the conventional QFT result. We have  
\begin{equation}
\Delta(g-2) \sim \frac{\alpha}{\pi} \left [\left (\frac{m_{\mu}}{\Lambda}
\right )^2 + \frac{1}{m_{\mu}^2 L^2} \right ]\;,
\end{equation}
where $m_{\mu}$ is the muon mass. Unlike in conventional QFT, this
contribution cannot be eliminated by taking $\Lambda, L \rightarrow \infty$,
because of the intrinsic relationship between the cutoffs,
\begin{equation}
N \Lambda^4 \lsim L^{-2}M_{Pl}^2 \;.
\end{equation}
Using (5) in (4) to eliminate $L$, and then minimizing the whole expression
with respect to $\Lambda$, one ends up with a minimal contribution beyond 
the conventional QFT which no way can be avoided,
\begin{equation}
\Delta(g-2, min) \gsim \; \frac{\alpha}{\pi} 
\left ( \frac{m_{\mu}}{M_{PL}} \right )^{2/3} N^{1/3}
\;.
\end{equation}
From the report of the muon E821 anomalous magnetic moment measurement at
BNL \cite{13} we know that 
\begin{equation}
\frac{g_{\mu}-2}{2}(exp - SM) = (22 -26) \times 10^{-10} \;.
\end{equation}
In turn, this implies
\begin{equation}
\left ( \frac{m_{\mu}}{M_{PL}} \right )^{2/3} N^{1/3} \lsim 10^{-7} \;,
\end{equation}
leading to a very strong bound
\begin{equation}
N \lsim 10^{19} \;.
\end{equation}
On the other hand, the upper bound (3), relevant for populating the the
above-mentioned gap in entropy, can be estimated for $\Lambda$
minimizing (4), $\Lambda_{min} \simeq N^{-1/6} m_{\mu}^{2/3} M_{Pl}^{1/3}$,
to be $\gsim \, 10^{47}$. Hence for such configurations, the bound on $N$
coming from high precision experiments would preclude completing the gap in
entropy between normal systems and black holes of the same size.

It might be instructive to compare the contribution of $N \sim 10^{19}$
particle species to the muon $g-2$ in a QFT endowed with the peculiar UV/IR
mixing of the type (5), with the the conventional Planck-scale correction
$(m_{\mu}/M_{Pl})^2 \sim 10^{-40}$, and with a correction 
arising from
$10^{32}$ species in a QFT not obeying the holographic bounds 
$(m_{\mu}/1 {\rm TeV})^2 \sim 10^{-8}$ (or
equivalently from $10^{32}$ KK particles in a higher-dimensional setting).
While the Planck-scale correction is vanishingly insignificant, the
contribution of $10^{32}$ species from a conventional QFT is rather
significant but still less than the contribution from a holographic theory
with $10^{19}$ species. This clearly demonstrates how in a holographic
theory 
one can extract useful information on $N$ from high precision
experiments much more efficiently than in a
conventional theory, where the bound on $N$ is obtained by noting that we
have not seen any strong gravity in the particle collisions.

Next we proceed with  an effect attributed to the relevant
operators ($D >4$) in considering processes at low energies $E$. The effect
explicitly depends on $E$ and a correction induced by these operators due
to the finite UV/IR cutoffs now reads,
\begin{equation}
f(\Lambda,L,E) \sim  \frac{\alpha}{\pi} 
\left [\left (\frac{E}{\Lambda} \right )^{D-4} + \frac{1}{L^2 E^2} \right ]
\;. 
\end{equation}
Again, using (5) and minimizing with respect to $\Lambda$ yields
\begin{equation}
\Lambda_{min} = [(D-4)/4]^{1/D} ~ E^{\frac{D-2}{D}} ~ M_{Pl}^{\frac{2}{D}}
~ N^{-\frac{1}{D}} \;.
\end{equation}
The internal consistency then requires
\begin{equation}
E \lsim \Lambda_{min} \;,
\end{equation}
leading to an upper bound on $N$
\begin{equation}
N \lsim \left (\frac{M_{Pl}}{E} \right )^2 \;.
\end{equation} 
Similarly, 
\begin{equation}
E \gsim \; L_{min}^{-1} \;,
\end{equation}
where $L_{min}^{-1} \gsim \; N^{1/2} \Lambda_{min}^2 M_{Pl}^{-1}$, precisely fills the gap from below,
\begin{equation}
N \gsim \; \left (\frac{M_{Pl}}{E} \right )^2 \;.
\end{equation}
The highest-energy particle collisions with $E \sim 1$ TeV
would then minimize the bound, and we see that a familiar value $N \sim 10^{32}$
is singled out. Notice that the bound is independent on the dimension $D$.
One may wonder how this bound comes out in a different theory using
unrelated arguments. The answer lies in the fact that $L_{min}^2 M_{Pl}^2$ 
always
(for any $D$) is of order $\sim M_{Pl}^2/E^2$, and that we have actually  
approached 
the black-hole limit where the two theories coincide. The 
internal-consistency arguments alone would always blow $N$ in (3) up to a 
saturation, approving the familiar bound.  

Now we use the holographic restriction (5) to constrain the  parameters
of the renormalization-group (RG) evolution in a conventional
field-theoretical model in curved space, i.e.,
\begin{equation}
\rho_{\Lambda }(\mu ) \; \simeq \; \mu^{2} \; G_{N}^{-1}(\mu ) \;,
\end{equation}
where $\rho_{\Lambda }$ is the zero-point energy density (substituting
$\Lambda^4$), $G_{N}^{-1}=M_{Pl}^2$ and  
we have identified the IR cutoff with the RG scale $\mu$. We have also
generalized the original relation (5) to incorporate the running character
of the Planck mass as well. We want to set constraints on a theory with a
large number $N$ of species with a mass scale $m$.

When the RG running scale $\mu $ is below the
lowest
mass in the theory, we can write the RG laws for  $\rho_{\Lambda }$ and
$G_N $ as \cite{14}
\begin{equation}
\rho_{\Lambda }(\mu) = \sum^{\infty}_{n=0}C_n \mu^{2n} \;,
\end{equation}
\begin{equation}
G_{N}^{-1}(\mu) = \sum^{\infty}_{n=0}D_n \mu^{2n} \;.
\end{equation}
We assume that both series
converge well and can be well approximated by retaining just a first few
terms. From the studies of the cosmologies with the running $\rho_{\Lambda
}$ and $G_{N}$ in the formalism of QFT in curved spacetime \cite{14, 15}
we
know that generally $C_1 \sim N m^2 $, $C_2 \sim N$, $C_3
\sim 1/m_{min}^2 $, etc.; $D_0 = M_{Pl}^2 $, $D_1 \sim N$, $D_2 \sim
1/m_{min}^2$ etc., where $m_{min}$ denotes the smallest mass of all massive
fields in the theory. Plugging (17) and (18) into (16) we get \footnote{We note
that from the holographic perspective $C_0$, representing the ground state 
of the vacuum, is always set to zero.} $C_n \simeq D_{n-1}$ and thus
\begin{equation}
N m^2 \simeq M_{Pl}^2 \;.
\end{equation}
If furthermore we formally equalize 
\begin{equation}
N m^2 \mu^2 \simeq N \Lambda^4 \;,
\end{equation}   
one sees that the black-hole limit $\mu \simeq \Lambda$ implies $m \simeq
\Lambda$, and again $\Lambda \simeq M_{Pl}/\sqrt{N}$ runs out. Much more
interesting is the other possibility allowed by (19) and (20), $m \gsim \;
\Lambda$. In this case
\begin{equation}
N \simeq \frac{M_{Pl}^2}{m^2} \lsim \frac{M_{Pl}^2}{\Lambda^2} \lsim
\frac{M_{Pl}^2}{\mu^2} \;,
\end{equation}
and the constraint of the type (3) now translates into much stronger one
involving the UV cutoff. We see that the allowable number of species $N$
gets progressively reduced as $m$ is elevated. Since we know that the mass
spectrum of elementary particles could expand up till  $\sim M_{Pl}$, the
limit on the number of species can be as low as $\sim 1$ for hugely massive
particles. It is curious to see how holography tends to expand the particle
spectrum towards $M_{Pl}$, at the same time switching  the role of the UV
cutoff  and  the particle mass.

In conclusion, we have considered the large-N limit of the four-dimensional
effective QFT obeying the holographic entropy bounds. As such it can thus
serve as a bridge from particle theory into quantum gravity. The bound on
the number of species from the muon $g-2$ has appeared so severe, that  
normal systems are prevented to evolve into black holes of the same size 
by increasing $N$. On the other hand, the internal consistency of the theory
seems to always pinpoint the benchmark value $N \simeq 10^{32}$ from the alike
theories. Finally, using the holographic entropy bounds to set constraints
on the RG evolution of both the cosmological constant and Newton's constant,
we see that the particle mass is to assume the role of the UV cutoff.

{\bf Acknowledgment. } This work was supported by the Ministry of Science,
Education and Sport
of the Republic of Croatia under contract No. 098-0982887-2872.

\end{document}